\documentstyle[aps,prl,twocolumn,epsf]{revtex}
\def\bfnabla{\mbox{\boldmath $\nabla$}}

\begin{document}
\twocolumn[\hsize\textwidth\columnwidth\hsize\csname
  @twocolumnfalse\endcsname
\title{\bf The infrared behaviour of the static potential in perturbative QCD}
\author {Nora Brambilla$^1$, Antonio Pineda$^2$, Joan Soto$^3$ and Antonio Vairo$^4$}
\address{$^1$ Institut f\"ur Theoretische Physik, 
     Boltzmanngasse 5, A-1090 Vienna, Austria}
\address{$^2$ Theory Division CERN, 1211 Geneva 23, Switzerland}
\address{$^3$ Dept. d'Estructura i Constituents de la Mat\`eria and IFAE, 
     U. Barcelona \\ Diagonal 647, E-08028 Barcelona, Catalonia, Spain}
\address{$^4$ Institut  f\"ur Hochenergiephysik, \"Osterr. Akademie d. Wissenschaften\\
     Nikolsdorfergasse 18, A-1050 Vienna, Austria}
\maketitle

\begin{abstract}
\noindent
The definition of the quark-antiquark static potential is given within an effective field theory framework. 
The leading infrared divergences of the static singlet potential in perturbation theory are explicitly 
calculated. \vspace{2mm}  
\end{abstract}

\vskip 2pc] 

\section{Introduction}

The static singlet potential between a heavy quark and an antiquark is an  
object of considerable interest both theoretically and phenomenologically.
Strictly speaking, being the potential a dynamical quantity, it can be defined 
only in a context where the full dynamics of the system is consistently taken into account. 
For instance in the proper Schr\"odinger equation.  
Nevertheless, the static singlet potential is usually defined as the 
logarithm of the 
static Wilson loop divided by the interaction time in the limit of infinite 
interaction time. Moreover, the above definition, in QCD, suffers from 
infrared (IR) divergences when computed at finite orders of perturbation 
theory \cite{Appelquist}. 
This has to do with the non-Abelian nature of QCD, 
which allows massless gluons to self-interact at arbitrarily small energy scales.

The problem was posed in the past \cite{Appelquist} of whether the static singlet 
potential could be defined in some way at any order in perturbation theory. 
The leading IR singularities in the static Wilson loop were found at a relative order 
$\alpha_{\rm s}^3$ in \cite{Appelquist}. These singularities can indeed be regulated upon resummation 
of a certain class of diagrams which give rise to a dynamical cut-off provided by the difference 
between the singlet and the octet potentials. However, the extraction of the static singlet 
quark-antiquark ($Q$-$\bar Q$) potential from this resummation has been regarded as suspect 
by other authors \cite{Brown}, who noticed that the difference between the singlet and the 
octet potential provides a dynamical scale which, at least in perturbation theory, is of the same order 
of the kinetic energy for quarks of large but finite mass. A similar situation occurs in QED for the Lamb shift, 
where one of the contributions has to be interpreted as a non-potential correction to the 
hydrogen atom energy levels. The novel feature in QCD is that such an effect appears 
to be related to the definition of the static potential, while in QED it is of order $1/m^2$. 
No quantitative procedure has been developed so far to deal with this problem in QCD. 

The recent complete calculation of the static potential at two loops 
\cite{twoloop} and its foreseen applications to top pair production at the 
Next Linear Collider \cite{Beneke} brings this problem to the edge of 
experimentally testable physics.
We feel that a clear resolution is urgently needed.

It is the aim of this paper to fully clarify this problem by providing a quantitative 
framework where the different scales that characterize a non-relativistic system can be properly 
taken into account. This is achieved by using an effective field theory approach. 

We will assume that $mv \gg \Lambda_{\rm QCD}$, $m$ and $v$ being 
the heavy-quark mass and velocity respectively ($v\ll 1$). Then we will systematically integrate out 
the hard ($\sim m$) and soft ($\sim mv$) degrees of freedom by performing in 
both cases a perturbative 
matching to suitable effective field theories. The integration of the hard scale gives rise to non-relativistic 
QCD (NRQCD) \cite{NRQCD}, whereas the integration of the soft scale gives rise to potential 
NRQCD (pNRQCD) \cite{pNRQCD}. In the latter effective theory only ultrasoft (US) degrees of freedom are 
left, which in this letter means the ones with energy much smaller than $mv$. 
The static potential can be understood as a matching coefficient in pNRQCD.

The main result of this work is to state rigorously what in perturbative QCD 
has to be understood as the $Q$-$\bar Q$ static potential, namely the relevant object
for the dynamics of $Q$-$\bar Q$ pairs with large but finite mass.   
This does not simply coincide with the static Wilson loop as usually computed, 
since this turns out to contain US contributions as well. As a consequence 
the static potential manifests, at the relative order $\alpha_{\rm s}^3$, an explicit dependence 
on the cut-off of the effective theory and has to be understood as a matching coefficient in pNRQCD. 
The leading cut-off dependence is evaluated explicitly.

\section{Theoretical Framework}

After integrating out the hard scale ($\sim m$) from QCD, we are left with 
NRQCD. For 
our purposes it is sufficient to work at the lowest order in the NRQCD Lagrangian, namely, 
\begin{eqnarray}
{\cal L}_{\rm NRQCD} \! &=& \! \psi^\dagger \! \left\{ i D^0 + {{\bf D}^2\over 2 m} \right\} \!\psi +
\chi^\dagger \! \left\{ i D^0 - {{\bf D}^2\over 2 m} \right\} \!\chi \nonumber \\
& & - {1\over 4} F_{\mu \nu}^{a} F^{\mu \nu \, a}\,,
\label{lagnrqcd}
\end{eqnarray}
where $\psi$ is the Pauli spinor field that annihilates the fermion and $\chi$
is the Pauli spinor field that creates the antifermion; $i D^0=i\partial_0 -gA^0$ 
and $i{\bf D}=i{\bfnabla}+g{\bf A}$. 

Integrating out also the soft scale, $m v$, from (\ref{lagnrqcd}) we are left 
with an effective theory (pNRQCD) where only US degrees of freedom remain 
dynamical. The surviving fields are the $Q$-$\bar Q$ states  (with US energy) 
and the US gluons.  The $Q$-$\bar Q$ states can be decomposed into a singlet (S) 
and an octet (O) under colour transformation. The relative coordinate ${\bf r}= {\bf x}_1-{\bf x}_2$, 
whose typical size is the inverse of the soft scale, is explicit and can be considered as small 
with respect to the remaining (US) dynamical lengths in the system. Hence the gluon fields
can be systematically expanded in $\bf r$ (multipole expansion). Therefore the pNRQCD Lagrangian 
is constructed not only order by order in $1/m$, but also order by order in 
${\bf r}$. As a typical feature of an effective theory, all the non-analytic 
behaviour in ${\bf r}$ is encoded in the matching coefficients, which can be interpreted as potential-like terms. 

The most general pNRQCD Lagrangian density that can be constructed with these 
fields and that is compatible with the symmetries of NRQCD is given at the leading order in the multipole 
expansion by:
\begin{eqnarray}
& & {\cal L}_{\rm pNRQCD} =
{\rm Tr} \Biggl\{ {\rm S}^\dagger \left( i\partial_0 - {{\bf p}^2\over m} 
- V_s(r) + \dots  \right) {\rm S} 
\nonumber \\
& &\qquad + {\rm O}^\dagger \left( iD_0 - {{\bf p}^2\over m} 
- V_o(r) + \dots  \right) {\rm O} \Biggr\}
\nonumber\\
& &\qquad + g V_A ( r) {\rm Tr} \left\{  {\rm O}^\dagger {\bf r} \cdot {\bf E} \,{\rm S}
+ {\rm S}^\dagger {\bf r} \cdot {\bf E} \,{\rm O} \right\} 
\nonumber \\
& &\qquad  + g {V_B (r) \over 2} {\rm Tr} \left\{  {\rm O}^\dagger {\bf r} \cdot {\bf E} \, {\rm O} 
+ {\rm O}^\dagger {\rm O} {\bf r} \cdot {\bf E}  \right\},  
\label{pnrqcd0}
\end{eqnarray}
where ${\bf R} \equiv ({\bf x}_1+{\bf x}_2)/2$, ${\rm S} = {\rm S}({\bf r},{\bf R},t)$ and 
${\rm O} = {\rm O}({\bf r},{\bf R},t)$ are the singlet and octet wave functions respectively. 
All the gauge fields in Eq. (\ref {pnrqcd0}) are evaluated 
in ${\bf R}$ and $t$. In particular ${\bf E} \equiv {\bf E}({\bf R},t)$ and 
$iD_0 {\rm O} \equiv i \partial_0 {\rm O} - g [A_0({\bf R},t),{\rm O}]$. 
$V_s$ and $V_o$ are the singlet and octet heavy $Q$-$\bar Q$ static potential respectively. 
Higher-order potentials in the $1/m$ expansion and the centre-of-mass kinetic 
terms are irrelevant here and are neglected. We define 
\begin{eqnarray}
V_s(r) &\equiv&  - C_F {\alpha_{V_s}(r) \over r}, 
\label{defpot}\\ 
V_o(r) &\equiv&  \left({C_A\over 2} -C_F\right) {\alpha_{V_o}(r) \over r}.
\nonumber
\end{eqnarray}
$V_A$ and $V_B$ are the matching coefficients associated in the Lagrangian (\ref{pnrqcd0}) 
to the leading corrections in the multipole expansion. Both the potentials and the coefficients 
$V_A$ and $V_B$ have to be determined by matching pNRQCD with NRQCD at a scale $\mu$ smaller 
than $m v$ and larger than the US scales. Since, in particular, $\mu$ is 
larger than $\Lambda_{\rm QCD}$ the matching can be done perturbatively. At the lowest order in the coupling 
constant we get $\alpha_{V_s} = \alpha_{V_o} = \alpha_{\rm s}$, $V_A=V_B=1$. 
In order to have the proper free-field normalization in the colour space we define 
\begin{equation}
{\rm S} \equiv { 1\!\!{\rm l}_c \over \sqrt{N_c}} S \quad \quad {\rm O} \equiv  { T^a \over \sqrt{T_F}}O^a, 
\label{norm}
\end{equation}
where $T_F=1/2$.

\section{Matching}

In this section we discuss how to perform the matching between NRQCD and 
pNRQCD. In particular we will concentrate on the effects produced by the 
leading corrections coming from the multipole expansion. 
 
The matching is in general done by comparing 2-fermion Green functions 
(plus external gluons at the ultrasoft scale) in NRQCD and pNRQCD, order by 
order in $1/m$ and order by order in the multipole expansion. If the soft 
scale is in the perturbative region of QCD (i.e. larger than 
$\Lambda_{\rm QCD}$), this can be done explicitly order by order in the 
coupling constant. If not, one can still perform the matching by subtracting the 
calculation in pNRQCD to the desired order of accuracy. The remaining term 
only contains the soft scale 
(up to US higher-order corrections) and goes in the pNRQCD Lagrangian as a new potential term. 

The matching can be done once the interpolating fields for $S$ and $O^{a}$ have been identified in NRQCD. 
The former need to have the same quantum numbers and the same 
transformation properties as the latter. We choose the following definitions for the singlet
$$
\chi^\dagger({\bf x}_2,t) \phi({\bf x}_2,{\bf x}_1,t) \psi({\bf x}_1,t)
=  Z^{1/2}_s(r) S({\bf R},{\bf r},t)
$$
and  for the octet 
\begin{eqnarray*}
& &\!\!\!\chi^\dagger({\bf x}_2,t) \phi({\bf x}_2,{\bf R},t)T^a 
\phi({\bf R},{\bf x}_1,t) \psi({\bf x}_1,t)
\\
& &\qquad\quad =Z^{1/2}_o(r) O^a({\bf R},{\bf r},t), 
\end{eqnarray*}
where 
$$
\phi({\bf y},{\bf x},t)\equiv {\rm P} \exp \Big\{ \! ig \displaystyle 
\int_0^1 \!\! ds \, ({\bf y} - {\bf x}) \cdot {\bf A}({\bf x} - s({\bf x} - {\bf y}),t) \!\Big\}.
$$ 
From the normalization condition (\ref{norm}) it follows, at the tree level, that $Z_s = N_c$ and $Z_o=T_F$.

In order to get the singlet potential, we choose the following Green function: 
\begin{eqnarray}
I &\equiv& \langle 0 \vert  \chi^\dagger(x_2) \phi(x_2,x_1) \psi(x_1) \nonumber\\ 
& & \qquad \times \psi^\dagger(y_1)\phi(y_1,y_2) \chi(y_2) \vert 0 \rangle. 
\label{vsmatch}
\end{eqnarray}
In NRQCD we obtain, at order $(1/m)^0$:
\begin{equation}
I = \delta^3({\bf x}_1 - {\bf y}_1) \delta^3({\bf x}_2 - {\bf y}_2) \langle W_\Box \rangle , 
\label{vsnrqcd}
\end{equation}
where $W_\Box$ is the rectangular Wilson loop with edges $x_1 = (T/2,{\bf r}/2)$, $x_2 = (T/2,-{\bf r}/2)$, 
$y_1 = (-T/2,{\bf r}/2)$ and  $y_2 = (-T/2,-{\bf r}/2)$. 
The symbol $\langle ~~ \rangle$ means the average over the gauge fields.

\begin{figure}
\makebox[0cm]{\phantom b}
\epsfxsize=7.95truecm \epsfbox{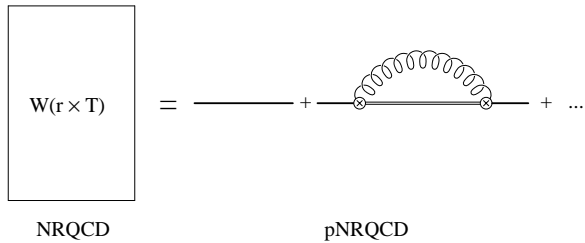}
\vspace{0.1cm}
\caption{The matching of the static potential. On the right side are the pNRQCD fields: 
simple lines are singlet propagator, double lines are octet propagators, 
circled-crosses  are the singlet-octet vertices of Eq. (\ref{pnrqcd0}) and 
the wavy line is the US gluon propagator.}
\label{figmat}
\end{figure}

In pNRQCD we obtain at order $(1/m)^0$ and at the  next-to-leading order in the 
multipole expansion
\begin{eqnarray}
& & 
I = Z_s(r) \delta^3({\bf x}_1 - {\bf y}_1) \delta^3({\bf x}_2 - {\bf y}_2) 
e^{-iTV_s(r)}
\nonumber\\
& & 
\times \Biggl(1-{ g^2 \over N_c} V_A^2 (r)\int_{-T/2}^{T/2} \!\! dt 
\int_{-T/2}^{t} \!\!dt^\prime e^{-i(t-t^\prime)(V_o-V_s)} 
\nonumber\\
& &
\quad \times\langle {\rm Tr} \{ {\bf r}\cdot {\bf E}(t) \phi(t,t^\prime)
{\bf r}\cdot {\bf E}(t^\prime) \phi(t^\prime,t)\}\rangle  \Biggr).
\label{vspnrqcdus}
\end{eqnarray}
Fields with only temporal argument are evaluated in the centre-of-mass coordinate.

\begin{figure}
\makebox[0.5cm]{\phantom b}
\epsfxsize=6truecm \epsfbox{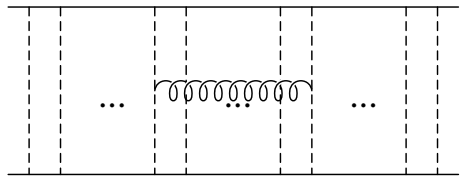}
\vspace{0.2cm}
\caption{Graphs contributing to $\delta \langle W_\Box \rangle$. Dashed lines represent Coulomb exchanges.}
\label{figapp}
\end{figure}

Comparing Eqs. (\ref{vsnrqcd}) and (\ref{vspnrqcdus}), one gets at the 
next-to-leading order in the multipole expansion the singlet wave-function normalization $Z_s$ 
and the singlet static potential $V_s$ (see Fig. \ref{figmat}). $V_A(r)$ and $V_o(r)$ must have 
been previously obtained from the matching of suitable operators. Since here we shall only 
need the tree-level results, we postpone the general discussion to \cite{BPSV1}.
We shall further concentrate on the singlet potential while the wave function normalization will 
also be discussed in \cite{BPSV1}. The explicit calculation can be done in several (equivalent) ways:

i) By matching order by order in $\alpha_{\rm s}$ \cite{BPSV1}. Both expressions (\ref{vsnrqcd}) and 
(\ref{vspnrqcdus}) are IR-divergent (Eq. (\ref{vspnrqcdus}) is also UV-divergent). Regulating both expressions in 
dimensional regularization the calculation in pNRQCD gives zero (there is no scale), while the calculation 
of the Wilson loop shows up an explicit dependence on the infrared regulator $\mu$ via a typical $\ln \mu r$ term.

ii) By keeping, without expanding, in (\ref{vspnrqcdus}) the exponentials in $V_o$ and $V_s$. 
The scale $V_o - V_s$, which appears in this way, regulates the IR divergences in 
Eq. (\ref{vspnrqcdus}). Because of this scale, the calculation in pNRQCD 
now gives a non-zero contribution proportional to the UV cut-off $\mu$ of 
the theory ($\sim \ln (V_o-V_s)/\mu$).  This calculation would be sufficient 
to extract the leading IR divergences of the static potential, in the same way as it is sufficient 
to evaluate the leading logarithm in the effective field theory in order to know the running of a matching 
coefficient. However, in order to carry out the matching consistently, we have to evaluate (\ref{vsnrqcd}) 
in a way that exactly corresponds to keeping the exponentials in (\ref{vspnrqcdus}). 
For the IR-singular terms in NRQCD, this way is nothing but the resummation of the diagrams depicted 
in Fig. \ref{figapp}, as it was indicated in \cite{Appelquist}.  
This procedure is automatically  adopted in any attempt to extract the (non-perturbative) 
static potential from a non-perturbative evaluation of the Wilson loop (e.g. in lattice calculations).

In order to make the comparison with \cite{Appelquist} easier, we will 
follow here this second approach. 
We will show the cancellation of the $\ln (V_o-V_s)$ terms 
explicitly. 
Evaluating expressions (\ref{vsnrqcd}) and (\ref{vspnrqcdus}) with the prescription ii) 
we get, for the static potential at the next-to-leading order in the multipole expansion:  
\begin{eqnarray}
&& V_s(r;\mu) = \lim_{T \rightarrow \infty}{i\over T} \ln {\langle W_\Box \rangle}
\nonumber \\
&& + C_F{\alpha_{\rm s} \over \pi} {r^2 \over 3}(V_o-V_s)^3 \ln {(V_o-V_s)^2 \over 4\pi\mu^2}.  
\label{pots0}
\end{eqnarray}
The dependence on $\ln (V_o-V_s)$ in the second line is canceled in the first line by the contribution 
to the Wilson loop ($\delta\langle W_\Box\rangle$) coming from the graphs shown in Fig. \ref{figapp} whose 
qualitative features were considered in \cite{Appelquist,kummer}. Indeed, the exact calculation done in this work 
gives 
\begin{eqnarray}
\lim_{T \rightarrow \infty}{i\over T} \ln {\delta \langle W_\Box \rangle} &=& 
- C_F C_A^2{\alpha_{\rm s}^3 \over 12 \pi} (V_o-V_s) 
\nonumber\\
& &\qquad \quad\times \ln \left({(V_o-V_s)^2 r^2}\right).  
\label{appdelta}
\end{eqnarray}

Finally, the result given in Eq. (\ref{pots0}) allows us to write the complete expression 
for the static potential, Eq. (\ref{defpot}), up to the relative order $\alpha_{\rm s}^3 \ln \mu r$ 
in coordinate space ($\alpha_{\rm s}$ is in the $\overline{\rm MS}$ scheme):
\begin{eqnarray}
&&{\alpha}_{V_s}(r, \mu)=\alpha_{\rm s}(r)
\left\{1+\left(a_1+ 2 {\gamma_E \beta_0}\right) {\alpha_{\rm s}(r) \over 4\pi}\right.
\nonumber\\
&&+\left[\gamma_E\left(4 a_1\beta_0+ 2{\beta_1}\right)+\left( {\pi^2 \over 3}+4 \gamma_E^2\right) 
{\beta_0^2}+a_2\right] {\alpha_{\rm s}^2(r) \over 16\,\pi^2}
\nonumber\\
&&\left. + {C_A^3 \over 12}{\alpha_{\rm s}^3(r) \over \pi} \ln{ r \mu}\right\},
\label{newpot}
\end{eqnarray}
where $\beta_n$ are the coefficients of the beta function, $a_1$ was 
calculated in Ref. \cite{1loop} and $a_2$ in Ref. \cite{twoloop} 
(see \cite{twoloop} for notation). We emphasize that this new contribution 
to the static potential would be zero in QED.

\section{Conclusions}

Eq. (\ref{newpot}) is the main result of this letter. It states that 
$\alpha_{V_s}$, defined through the static potential (see Eq. (\ref{defpot})), is not a short distance 
quantity as $\alpha_{\rm s}$ (in the ${\rm \overline{MS}}$ scheme), since it 
depends on the IR behaviour of the theory. 
It can be better understood as a matching coefficient (an analogous situation occurs for 
the pole mass). Moreover Eq. (\ref{newpot}) gives the analytic value of the coefficient 
of the leading $\ln \mu r$ correction, which arises at 
relative order $\alpha_{\rm s}^3$, i.e. immediately after the known two-loop corrections. 
The situation for the static octet potential seems to be similar \cite{BPSV1}. 

The evaluated terms clarify the long-standing issue of how the perturbative static potential 
should be defined at higher order in the perturbative series. Eq. (\ref{pots0}) explicitly shows 
that the static potential does not coincide with the static Wilson loop as usually 
computed. It should be emphasized that the separation between soft and US contributions is not an
artificial trick but a necessary procedure if one wants to use the static potential in a Schr\"odinger-like equation
in order to study the dynamics of $Q$-$\bar Q$  states of large but finite mass. In
that equation the kinetic term of the $Q$-$\bar Q$ system is US and so is the 
energy. Since the US gluons interact with the $Q$-$\bar Q$ system, their dynamics is sensitive
to the energies of the (non-static) system and hence it is not correct to include them in
the static potential. When calculating a physical observable the $\mu$ dependence in (10) must
cancel against $\mu$-dependent contributions coming from the US gluons. 

Finally it is worth mentioning that the static potential suffers from IR 
renormalons ambiguities with the following structure
$$
\delta V_s \sim C + C_2 r^2 + .\,.\,.
$$
The constant $C$ is known to be cancelled by the IR pole mass renormalon 
\cite{thesis} while the IR $C_2$ renormalon gets cancelled by the IR 
renormalon existing in the second term of Eq. (\ref{vspnrqcdus}) \cite{BPSV1}.

N.B. acknowledges the TMR contract No. ERBFMBICT961714, A.P. the TMR contract No. ERBFMBICT983405, 
J.S. the AEN98-031 (Spain) and 1998SGR 00026 (Catalonia) and A.V. the FWF contract No. 9013.

\end{document}